\documentclass[sn-vancouver]{sn-jnl}

\usepackage{amsmath}

\jyear{2021}%

\theoremstyle{thmstyleone}%
\theoremstyle{thmstyletwo}%

\theoremstyle{thmstylethree}%
\newcommand{\explorer}{United Imaging uEXPLORER\xspace}
\newcommand{\siemens}{Siemens Biograph Vision Quadra\xspace}
\begin{document}

\title[Parameterised Geant4 simulation for total body PET research]{Parameterised Geant4 simulation for total body PET research}


\author*[1]{\fnm{Benjamin M.} \sur{Wynne}}\email{b.m.wynne@ed.ac.uk}

\author[1]{\fnm{Hanna} \sur{Borecka-Bielska}}

\author[1]{\fnm{Matthew} \sur{Needham}}

\author[2,3]{\fnm{Adriana A. S.} \sur{Tavares}}

\affil[1]{\orgdiv{School of Physics and Astronomy}, \orgname{University of Edinburgh},
\orgaddress{\city{Edinburgh}, \country{UK}}}

\affil[2]{\orgdiv{BHF-University Centre for Cardiovascular Science}, \orgname{University of Edinburgh},
\orgaddress{\city{The Queen's Medical Research Institute, Edinburgh BioQuarter}, \country{UK}}}

\affil[3]{\orgdiv{Edinburgh Imaging}, \orgname{University of Edinburgh},
\orgaddress{\city{The Queen's Medical Research Institute, Edinburgh BioQuarter, Edinburgh}, \country{UK}}}




  \abstract{\textbf{Background:} Total-body positron emission tomography (PET) imaging has the potential to transform medical care of a number of diseases and augment our knowledge of systems biology. Various detector designs and geometries are currently under development for total\nobreakdash-body PET imaging of humans. This variety, in particular the variation in axial field-of-view (aFOV), motivates a need to compare the performance of these devices in a consistent simulated environment.

 \textbf{Methods:} We present an open-source Geant4 simulation package that allows variation of relevant parameters such as the detector aFOV and the tracer radioisotope from the command line.
 Two simplified detector geometries based on the \siemens and \explorer models are supported with variable granularity. 
 The intrinsic radioactivity of the detector crystals is fully simulated.
 The simulation can be viewed with the built-in GUI, and the results are saved in a plain text format for easy analysis.
 Example Python analysis code is provided with the simulation, demonstrating calculation of the noise equivalent count rate (NECR) figure of merit using an approximation to the NEMA NU 2-2012 standard method.

 \textbf{Results:} A good agreement between the simulated count rate performance and experimental data is observed for both geometries. The differences in results are attributed to simplifications in the simulation code, namely not accounting for the light-collection efficiency or readout dead-time. We demonstrate the importance of assessing the scanner performance using appropriate phantom length which significantly affects the obtained results. A dependence between the detector aFOV and the length of the source, with peak NECR plateauing as the detector extends beyond the region of interest is also presented.
 
 \textbf{Conclusion:} Simulation allows for the rapid assessment of detector performance in a variety of scenarios, demonstrating a link between aFOV and the test source length that was not explicit in published experimental data.
 Further studies with low source activity, or detector intrinsic radioactivity, may also demonstrate the importance of detector aFOV.}

\keywords{PET, Total-Body PET, Simulation, Geant4}



\maketitle

\section{Background}\label{sec:intro}

A conventional PET scanner has an axial field-of-view (aFOV, the length of the detector along the cylindrical $z$-axis) of around $20$\,cm.
More recently, `total-body' PET scanners, with aFOV $>1$\,m were proposed~\cite{cherryTBPET,StateTBPET}, and models with differing geometries and aFOV values are coming to market \cite{quadra,explorer}. A total-body PET scanner improves in several aspects compared to previous PET scanners. Its extended axial field of view covers nearly the entire body, allowing to make a single scan without adjusting bed position. This not only provides a more comprehensive insight into the body's metabolism and the distribution of radiotracers throughout the system but also significantly reduces scan time, enhancing patient comfort and minimising movement artefacts that might impact image quality.
The comprehensive perspective offered by total\nobreakdash-body PET scanners improves the assessment of how different parts of the body interact and influence each other. This proves especially valuable in pinpointing metastases, tracking disease progression, and devising treatment strategies across multiple body regions simultaneously.
A notable enhancement of total-body PET is the reduced radiation dose it offers. This decrease is beneficial for patients, reducing their exposure to radiation, which is of particular importance in paediatric medicine, while maintaining high-quality diagnostic images \cite{cherryTBPET}.

Assessment of promising research directions, or the consideration of the purchase of expensive new equipment~\cite{crystal-cost}, requires comparison of the performance of different approaches.
While simulated and experimental results are available, and National Electrical Manufacturers Association (NEMA) standards are defined for conventional PET scanner characterisation, such comparison can still be cumbersome. 
Several software packages exist that allow simulation of the transport of particles through matter. These encompass versatile tools such as Geant4~\cite{geant4}, EGS4~\cite{EGS4}, and MCNP~\cite{MCNP}. Within the domain of medical applications targeted towards PET research, solutions like GATE~\cite{GATE}, PETSIM~\cite{PETSIM-MC}, SIMSET~\cite{SIMSET}, PeneloPET~\cite{peneloPET}, and gPET~\cite{gPET} have been developed. In this article, we demonstrate a Geant4 simulation that allows easy variation of parameters such as aFOV, and provides a consistent environment for evaluating widely-varying technologies.
We compare simulated performance with published results for specific geometries and demonstrate more general results by scanning the parameter space.

\section{Methods}\label{sec:methods}

\subsection{Detector simulation}\label{sec:simulation}

\begin{figure}[ht]
\centering
\includegraphics[width=0.9\textwidth]{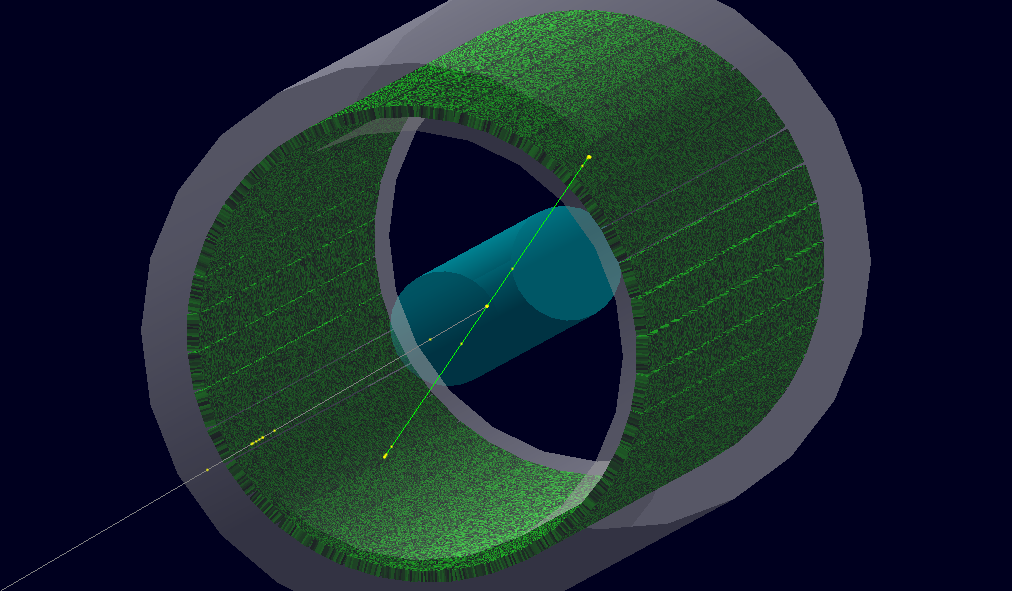}
\caption{Example ``true'' event from the Geant4 simulation, showing a pair of back-to-back photons (bright green lines) originating from a positron annihilation.
Although the neutrino from the nuclear decay will not be detected, it is still shown in the simulation as a grey line.
The cyan cylinder is the polyethylene phantom, and scintillator crystals are shown in dark green, with random variation in intensity so that their edges can be distinguished.
The grey tube enclosing the detector is for illustration only, and does not represent a physical object.}
\label{fig:TruePetPhotons}
\end{figure}

The PET scanner simulation is based on the Geant4~\cite{geant4} toolkit and the code is available for download~\cite{github-repo}.
A single unit of the scintillating material is defined, and then duplicated many times (potentially O$(10^5)$) at different positions.
This allows an entire PET detector to be created from a simple mathematical description of the scintillator placement, the parameters of which can then be varied without any corresponding code change.
In this way, two simplified detector geometries based on the \siemens~\cite{quadra} and \explorer~\cite{explorer} models are defined.
The smallest unit of scintillator crystal (as described by the manufacturers) is used for each detector, but versions are also available that combine multiple crystals into a single block, allowing a trade-off between simulation fidelity and speed.
At present, no attempt is made to simulate the performance of the detector electronics; it is assumed that energy deposited in the scintillator will be recorded.
Besides the scintillator crystals, the only other object included in the simulation is a cylindrical polyethylene phantom in the centre of the detector, as illustrated in Figure~\ref{fig:TruePetPhotons}. The default phantom has a diameter of 20.3~cm and a length of 70~cm. The tracer is uniformly distributed along the $z$-axis in the middle of the phantom.

The simulation can run in two modes: either modelling radioactive decay of a tracer isotope, or the intrinsic radioactivity of the detector itself.
Individual atoms from either source are allowed to decay using the built-in radioactivity models provided by Geant4.
Atoms of the tracer isotope are placed at a random position along the $z$-axis of the detector, \textit{i.e.} along the centre of the phantom. This is not exactly as defined in the NEMA NU 2 standard~\cite{NEMA-NU-2-2012}, which requires a linear source parallel to, but offset from, the $z$-axis.
However, this change simplifies the geometry of the system allowing for rapid calculation of figures-of-merit such as the noise equivalent count rate (NECR), without compromising the utility of the results.
When simulating the intrinsic radioactivity of the detector, atoms are placed at a random $z$ and $r$ coordinate within the scintillator material, but are constrained to $\phi=0$.
This takes advantage of the fact that detectors are assumed to be (approximately) symmetrical in $\phi$, as is the phantom.

The simulation can be run from the command line, specifying parameters such as the aFOV and detector material as arguments.
The output data from the simulated decays are saved as a plain text file for ease of use.
A 3D-rendering of the detector and simulated events is available, as provided by Geant4 \cite{geant4} and shown in Figure~\ref{fig:TruePetPhotons}.

\subsection{Analysis}\label{sec:analysis}

We provide example Python code in the form of Jupyter notebooks, which read the output files from the simulation and can reproduce the results in this communication.
The intention is for others to be able to use the simulation in their own analyses, and so these examples are by no means exhaustive or definitive.
Nonetheless, our approach is explained below.

PET detectors have characteristically large output datasets, beyond what is practical to simulate on a regular basis.
Hence, we use a relatively small (O$(10^6)$) collection of simulated decay events for a given detector geometry and randomly resample them over the course of an analysis.
When simulating a period of data-taking, radioactive decays are requested at time intervals following a simple Poissonian model, and an event from the simulation is associated with each one.
The length of the simulated data-taking period used is 10\,ms, again to reduce the resources required by the simulation. 
High and low energy thresholds defined as 435.0 and 585.0~keV for \siemens, and similiarly to 430.0 and 645.0~keV for \explorer, are applied for the detected photons, corresponding to recommended clinical values. A coincidence time-window of length $4.7$\,ns (consistent with Reference~\cite{quadra}) is opened if a photon passes the criteria. If at the end of the window exactly two photons have passed the energy requirements, this pair is analysed using a simplified technique.

\begin{figure}[h]
\centering
\includegraphics[width=0.8\textwidth]{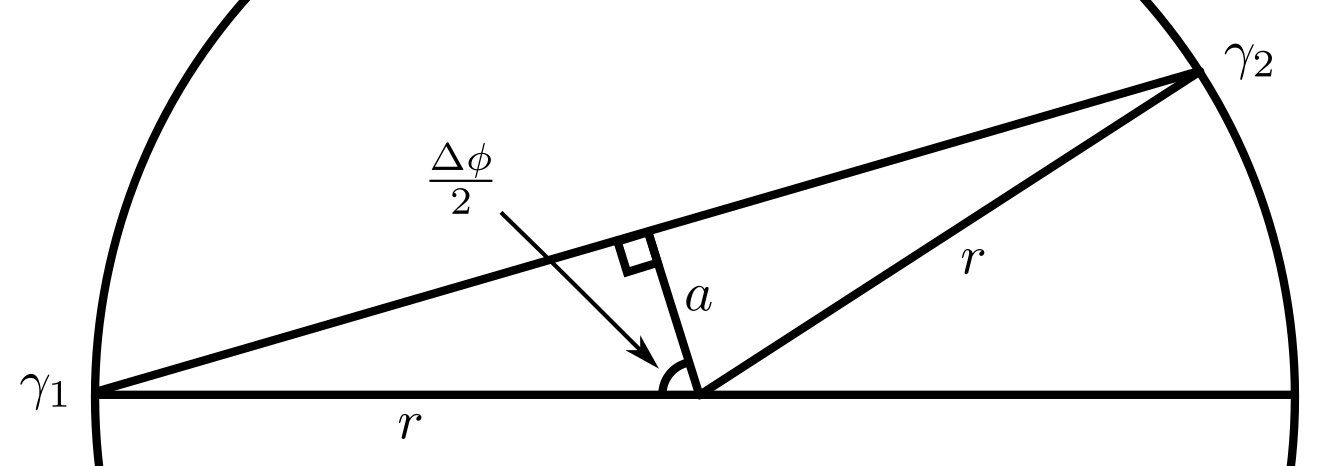}
\caption{Schematic cross-section of simulated detector, showing the geometrical relationship between photon detections $\gamma_1$ and $\gamma_2$, the difference in their $\phi$ coordinates $\Delta\phi$, and the minimum offset $a$ of the line of response from the centre of the detector of radius $r$. }
\label{fig:angles}
\end{figure}

Rather than constructing sinograms from the photon pairs, we take advantage of the earlier choice to simulate the tracer radioisotope as a linear source along the detector $z$-axis.
An approximation to the standard NECR value can be calculated using only $\Delta\phi$ values as input.
The offset of the line of response from the centre of the detector (i.e. the $z$-axis) is $a=r\cos{\frac{\Delta\phi}{2}}$, given the simple geometry shown in Figure~\ref{fig:angles}.

This offset, $a$, is analogous to the ``radial distance from maximum pixel [of the sinogram]'' used in the NEMA standard, since the line source is guaranteed to lie on the $z$-axis of our simulation.
We create a histogram $C(a)$ of the $a$ values, and events with $20<\lvert a \rvert<120$\,mm are taken to be background.
In the central region $\lvert a \rvert<20$\,mm, the contribution from background is calculated by linear interpolation between the values found at $a=\pm20$\,mm.
This contribution is subtracted from the total in the central region to give the true events.
Events with $\lvert a \rvert>120$\,mm are ignored.
The calculation is then re-run with the requirement that both detected photons originated from the same positron, using their identifiers from the simulation output.
This allows the (unphysical) elimination of random coincidences, and thus the scattered event rate can be calculated separately, and subtracted from the total background to give the random rate.
Thus the rates for all events ($R_{TOT}$), true ($R_{t}$), random ($R_{r}$), and scatter ($R_{s}$) events, and NECR ($R_{NEC}$) in a data acquisition time, $T_{acq}$, can be expressed as 
\begin{center}
\begin{equation*}
\begin{split}
R_{t} &= \frac{1}{T_{acq}} \sum_{\lvert a \rvert<20} \left( C(a) - \frac{1}{2} \Bigl( C(-20) + C(20) \Bigr) \right),\\
R_{TOT} &= \frac{1}{T_{acq}} \sum_{\lvert a \rvert<120} C(a),\quad R_{NEC} = \frac{R_{t}^{2}}{R_{TOT}},\\
R_{r+s} &= R_{TOT} - R_{t},\quad R_{s} = R'_{TOT} - R'_{t},\quad R_{r} = R_{r+s} - R_{s}.
\end{split}
\end{equation*}
\end{center}

Here $R_{r+s}$ is the combined background rate, and $R'$ denotes values calculated with the constraint that detected photons must originate from the same positron.
The histogram binning is chosen so that $a=\pm20~\textrm{mm}$ must be bin edges, and so the values $C(-20)$ and $C(20)$ are given by the mean of the counts in the adjacent bins.

\subsection{Uncertainty estimation}\label{sec:uncertainties}

Emphasising the speed and the simplicity of approach, most example results do not include an uncertainty estimation, only a single value for each data point.
However, three sources of experimental uncertainty are examined in a separate calculation.

The calculation of NECR at a given time-point requires the collection of a set of simulated photon pairs.
Consequently, there is a Poissonian statistical uncertainty on the NECR value arising from the total number of photon pairs collected.
Typically, the relative energy and time resolution values for current-generation PET scanners are around 10\% and $0.5$\,ns respectively.
The impact of these uncertainties is estimated by re-running calculations as a set of pseudoexperiments.
In each trial, the energy and detection-time values for the photons have random, Gaussian-distributed perturbations applied, with the Gaussian $\sigma$ parameter corresponding to the resolution value.
Sampling the simulated data and applying the resolution effects is controlled by an independent random number generator in each pseudoexperiment.
The uncertainty is then taken to be the envelope of the results from all pseudoexperiments, and the mean values are taken as nominal. Ten pseudoexperiments are used.

The effect of increasing the number of simulated events collected in a pseudoexperiment is to reduce all the uncertainties mentioned, at the cost of CPU time.
For the results presented here, the $10$\,ms simulated data taking periods give rise to uncertainties of approximately $\pm2$\,\%, and this value should be assumed unless others are explicitly stated.

\section{Results}\label{sec:results}

\begin{figure}[ht]
\centering
\begin{subfigure}{0.49\textwidth}
\includegraphics[width=\linewidth]{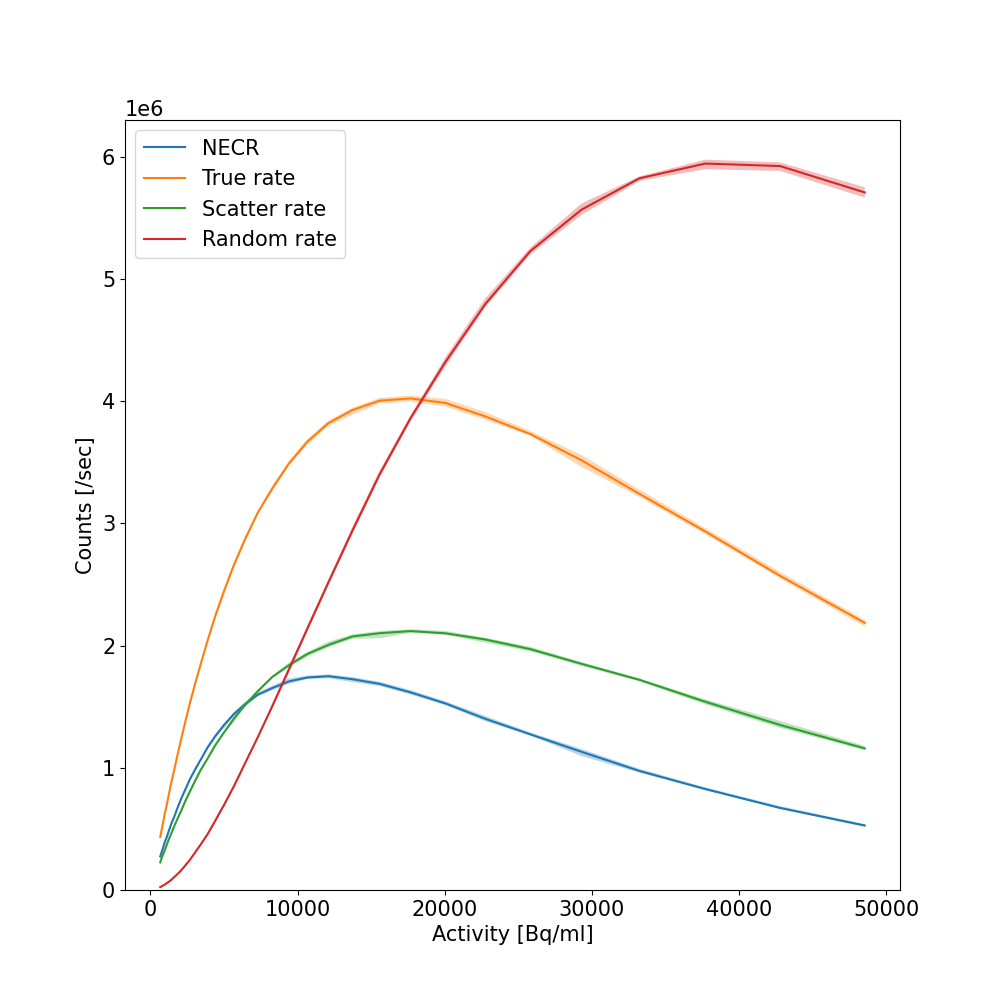}
\caption{\siemens} \label{fig:SiemensVsExplorer:siemens}
\end{subfigure}
\begin{subfigure}{0.49\textwidth}
\includegraphics[width=\linewidth]{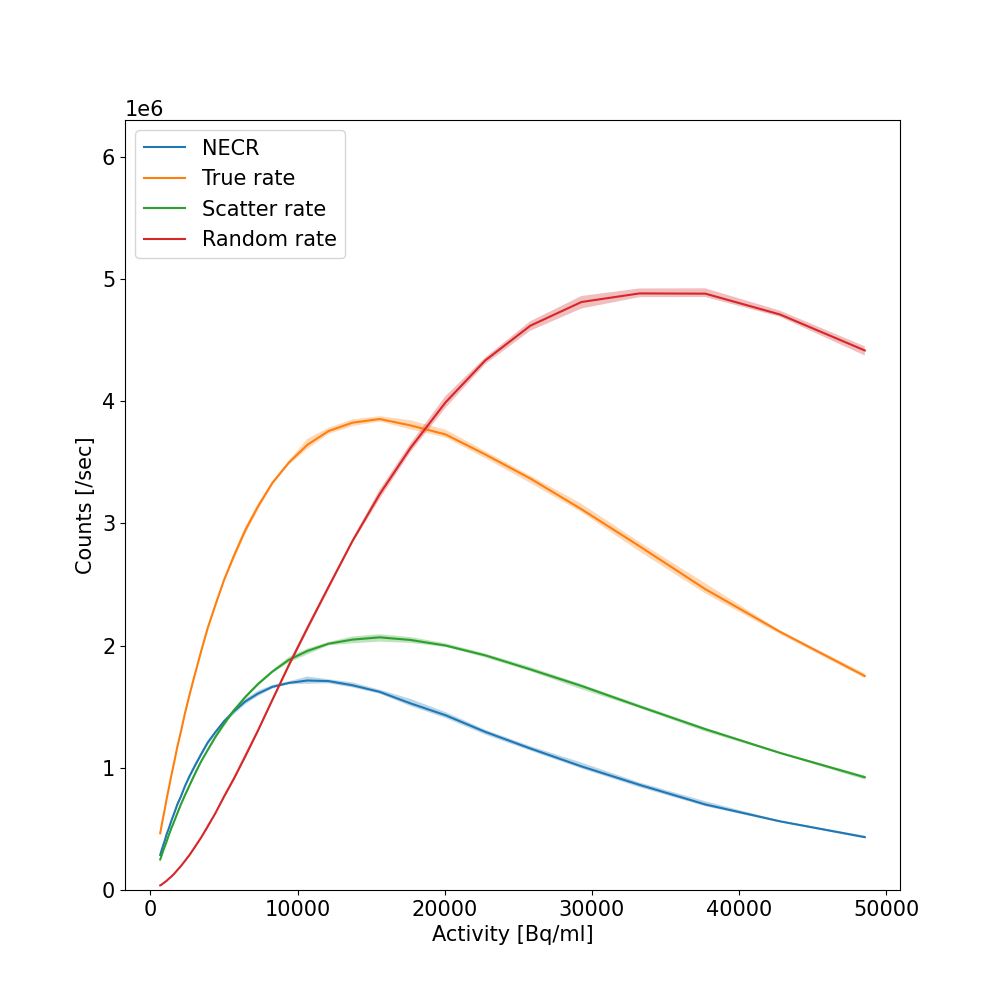}
\caption{\explorer} \label{fig:SiemensVsExplorer:explorer}
\end{subfigure}
\caption{NECR calculated with respect to source activity following the NEMA NU 2-2012 prescription presented for \siemens (\ref{fig:SiemensVsExplorer:siemens}) and \explorer (\ref{fig:SiemensVsExplorer:explorer}). True, scatter and random rates used in the calculation of the NECR are plotted separately. Shaded bands show the uncertainty range.}
\label{fig:SiemensVsExplorer}
\end{figure}

Figure~\ref{fig:SiemensVsExplorer} presents the NECR along with the true, scatter and random coincidence rates, calculated following the NEMA NU 2-2012 standard, for the two tested total-body PET geometries. The \siemens (Reference~\cite{quadra}, Figure~\ref{fig:SiemensVsExplorer:siemens}) 
and the \explorer (Reference~\cite{explorer}, Figure~\ref{fig:SiemensVsExplorer:explorer}) results show similar maximum NECR of $\sim1.5\times10^6$\,counts/sec at a source activity of $10$\,kBq/ml. The scatter fraction is stable across the whole activity range and for both geometries is on average 35\%, which agrees well with 37\% quoted for Siemens \cite{quadra} and United Imaging \cite{explorer} scanners.

\begin{figure}[ht]
\centering
\includegraphics[width=0.8\textwidth]{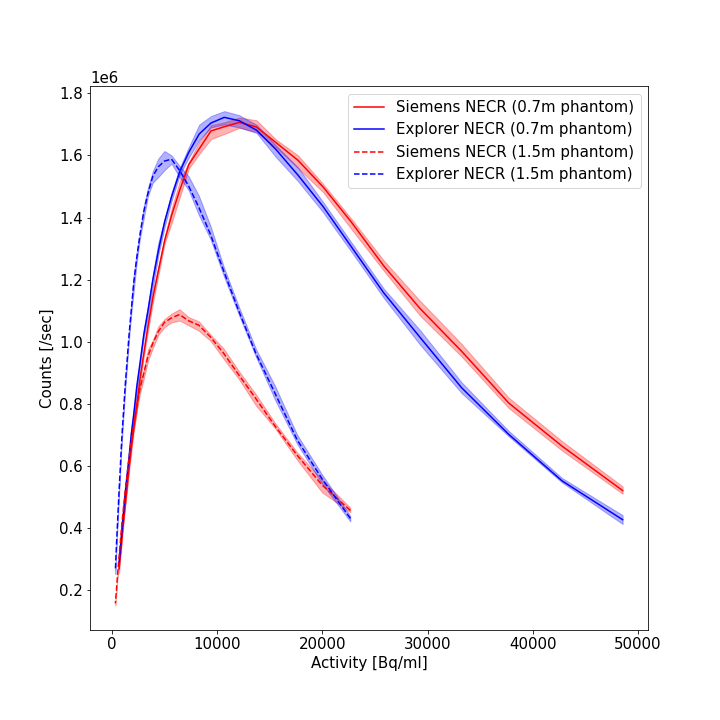}
\caption{NECR calculated with respect to source activity following the NEMA NU 2-2012 prescription presented for \siemens and \explorer with two phantom sizes: 70~cm and 150~cm axial length. 
Shaded bands show the uncertainty range.
}
\label{fig:SiemensVsExplorerTwoPhantoms}
\end{figure}
\FloatBarrier
Figure~\ref{fig:SiemensVsExplorerTwoPhantoms} shows a comparison when a 70~cm and $150$\,cm phantom is used to illustrate effect of phantom length on count rate performance. 

\begin{figure}[ht]
\centering
\begin{subfigure}{0.49\textwidth}
\includegraphics[width=\linewidth]
{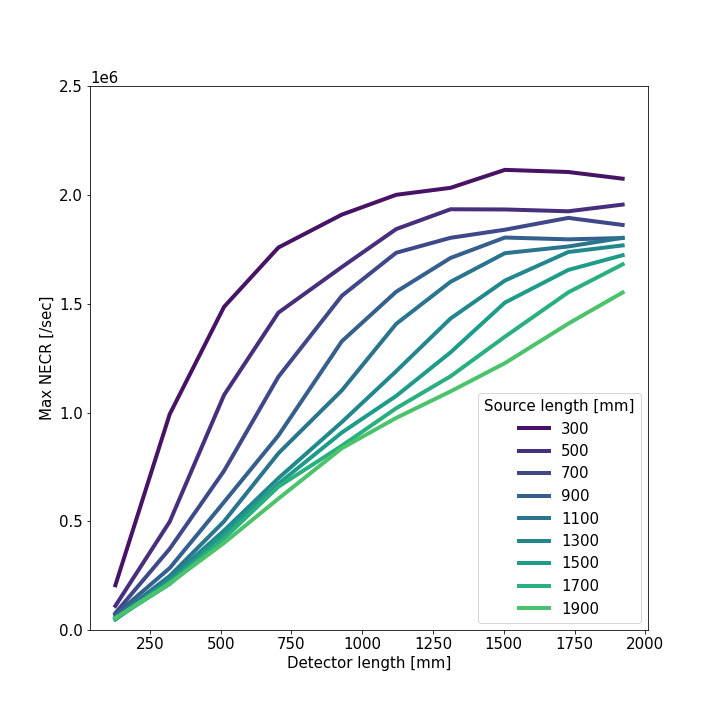}
\caption{} \label{fig:necr1100MBq:all}
\end{subfigure}
\begin{subfigure}{0.49\textwidth}
\includegraphics[width=\linewidth]
{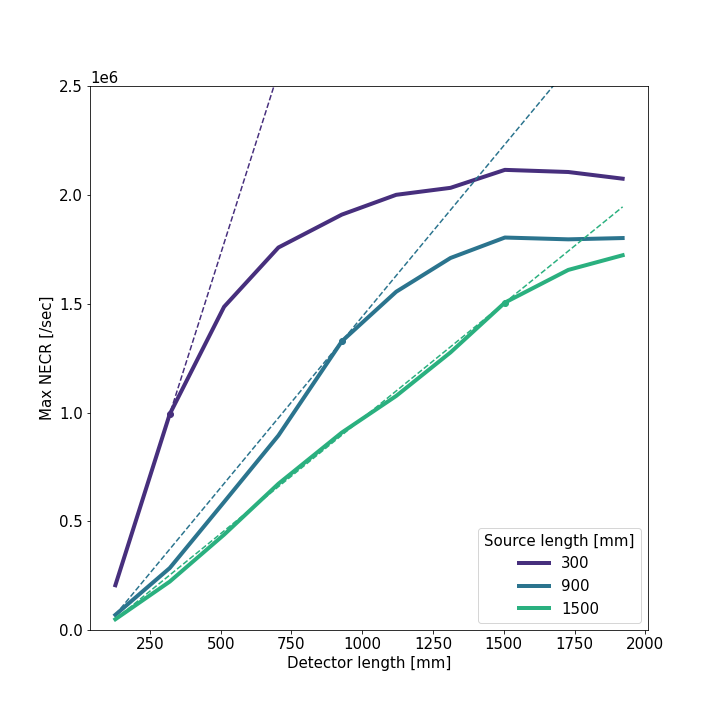}
\caption{} \label{fig:necr1100MBq:linear}
\end{subfigure}
\caption{Maximum NECR obtained for simulated variations in aFOV of the \siemens detector corresponding to different phantom and source lengths. Figure~\ref{fig:necr1100MBq:linear} shows a subset of the sources and includes additional guidelines to illustrate the initial linear NECR improvement. The results are presented for a source with initial activity of 1100~MBq.}
\label{fig:necr1100MBq}
\end{figure}
\begin{figure}[!ht]
\centering
\begin{subfigure}{0.49\textwidth}
\includegraphics[width=\linewidth]
{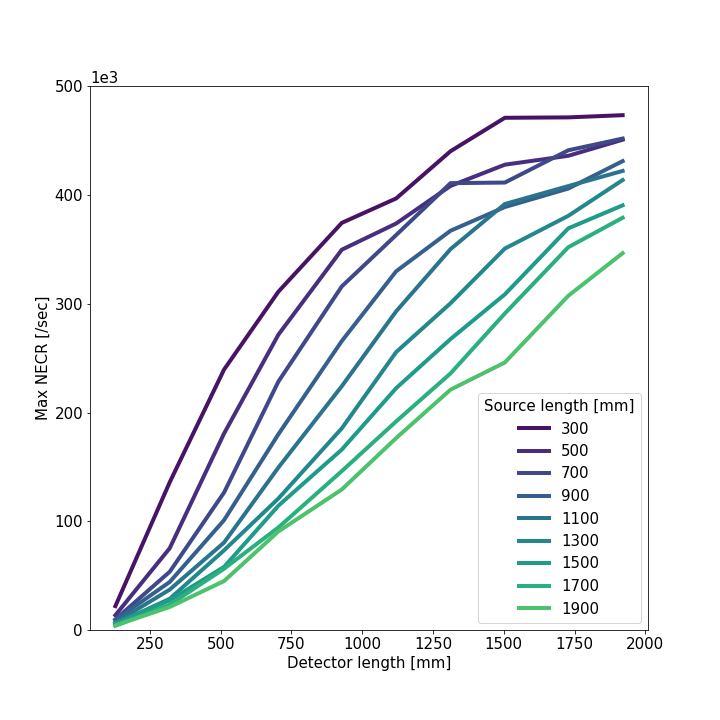}
\caption{} \label{fig:necr20MBq:all}
\end{subfigure}
\begin{subfigure}{0.49\textwidth}
\includegraphics[width=\linewidth]
{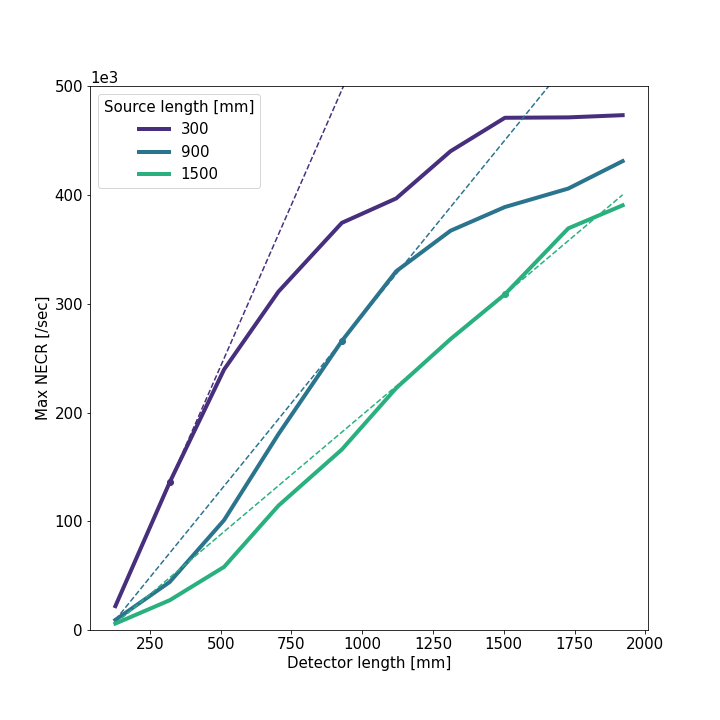}
\caption{} \label{fig:necr20MBq:linear}
\end{subfigure}
\caption{Maximum NECR obtained for simulated variations in aFOV of the \siemens detector corresponding to different phantom and source lengths. Figure~\ref{fig:necr20MBq:linear} shows a subset of the sources and includes additional guidelines to illustrate the initial linear NECR improvement. The initial source activity of $20$\,MBq is used to avoid results affected by detector saturation.}
\label{fig:necr20MBq}
\end{figure}
\FloatBarrier
Taking advantage of the parameterised geometry in the simulation, we repeat the study of NECR for a range of detector aFOV values and phantom lengths.
All results use aFOV variations of the \siemens detector --- while there are other distinctions between the \siemens and \explorer devices (\textit{e.g.} scintillator material) the choice of aFOV is most significant in this test.
For each simulation the starting activity is 1100~MBq following Reference~\cite{quadra}. The peak NECR value is obtained and compared in Figure~\ref{fig:necr1100MBq}. The same distributions produced with a starting source activity of 20~MBq to avoid a saturation effect are presented in Figure~\ref{fig:necr20MBq}.

\section{Discussion}\label{sec:discussion}

The trends of the simulated results shown in Figure~\ref{fig:SiemensVsExplorer} differ from experiment, peaking earlier and then declining at higher activity values. This is likely due to the simulation making no attempt to account for light collection efficiency or readout dead-time, thus displaying an idealised detector performance.
In the interests of allowing consistent comparison between detector geometries we do not attempt to reproduce more specialised behaviours of the systems, and thus such differences between simulated and experimental results will remain.

It is interesting to note that both detectors have similar performance, despite the \explorer aFOV being almost double that of the \siemens.
This can be explained by the NEMA NU 2-2012 standard phantom for NECR measurement being $70$\,cm in axial length.
Essentially, the phantom is of a similar size to the Biograph Vision Quadra detector, whereas the uEXPLORER detector is significantly longer --- length that could be considered wasted in this test. Distributions in Figure~\ref{fig:SiemensVsExplorerTwoPhantoms} demonstrate how important it is to define the phantom length to assess the scanner performance. The 150~cm long phantom extends well beyond the Siemens aFOV, but is still fully enclosed by the uEXPLORER detector. This leads to dramatically reduced performance from the shorter geometry.

We observe an approximately linear relationship between detector aFOV and maximum NECR as presented in Figure~\ref{fig:necr1100MBq} and Figure~\ref{fig:necr20MBq}, up to the point where the aFOV equals the phantom length.
As the detector becomes longer than the phantom, the improvement in performance slows and eventually plateaus.
For example, when measuring a $300$\,mm phantom, tripling the detector aFOV from $300$ to $900$\,mm only results in a doubling of NECR performance, and further extending the detector to $1900$\,mm gives minimal additional benefit.
This relationship is shown more clearly in Figure~\ref{fig:necr1100MBq:linear}. 
Note that each phantom has the same total activity, which explains why the shorter phantoms give the most rapid initial increase in detector performance with length: a greater percentage of the total activity is enclosed for a given increase in aFOV.
While increasing the detector aFOV is never demonstrated to be detrimental, the study shows that for a given size region of interest a similar length detector is most efficient.
Increasing length beyond the region of interest will improve performance at a limited rate, towards an upper bound.
The cost of current detector technology is largely driven by the scintillating crystals used~\cite{crystal-cost}. Extending the aFOV gives a corresponding increase in the amount of scintillator material and thus cost.
Once this is accounted for, it is clear that the ideal detector length may be less than the maximum available, depending on the intended use.
The potential of total-body PET scanners is to image a full adult body, and here a correspondingly long detector will be preferred.
However, if the intention is to study a smaller region of interest (\textit{e.g.} the torso, or a single organ) then a smaller detector may provide equivalent performance for lower cost.

This effect is not simply geometric: we have shown that it is related to detector saturation when the radiotracer activity is higher.
Different detectors may already account for this in the readout by imposing additional criteria on forming photon coincidences (\textit{e.g.} Reference~\cite{explorer}) but we do not simulate this.
Lower radiotracer activity also reduces potential saturation, although at the cost of the overall reduction in sensitivity that would be expected.
For a clinical procedure, less injected radioactivity would be beneficial for the patient, and we have shown that a longer detector aFOV does help recover lost sensitivity, although still with diminishing returns.

In general, rapid simulation approaches such as this allow optimisation of detector geometry for its intended role.
The space of a given parameter like aFOV can be uniformly scanned in a consistent environment, and ideal values identified, rather than attempting to compare different experimental results or develop specialised simulated models.

\section{Conclusion}\label{sec:conclusion}

Geant4 is used to simulate the \siemens and \explorer long-aFOV PET scan detectors, with additional parameters introduced to allow free variation of the detector geometry, scintillator material, or radioisotope.
The utility of this simulation is demonstrated by reproducing experimental results for the performance of these detectors, and then by freely varying the aFOV parameter with respect to the length of a phantom source.
We show that detector NECR improves linearly as aFOV is increased up to the length of the source, but that further improvement is less rapid and eventually plateaus.

The simulation package used for these studies is available for others to use and it is intended to develop it further, adding more detail and functionality. 

\backmatter

%
%

\section*{Declarations}

\subsection*{Ethics approval and consent to participate}
Not applicable
\subsection*{Consent for publication}
Not applicable
\subsection*{Availability of data and material}
All data used in the study was generated and analysed using the simulation code available at \url{https://github.com/bwynneHEP/SimplePetScanner}, and can be reproduced using this code.
\subsection*{Competing interests}
Not applicable
\subsection*{Funding}
This research is supported by a grant from the STFC cancer diagnosis network (CDN+, STFC Grant ST/S005404/1) held by BW. AAST was funded by the British Heart Foundation (FS/19/34/34354). AAST is a recipient of a Welcome Trust Technology Development Award (221295/Z/20/Z).
This publication has been made possible in part by CZI grant DAF2021-225273 and grant DOI https://doi.org/10.37921/690910twdfoo from the Chan Zuckerberg Initiative DAF, an advised fund of Silicon Valley Community Foundation (funder DOI 10.13039/100014989), to AAST.
\subsection*{Authors' contributions}
BMW wrote the simulation and analysis code, and was the principal contributor in writing the manuscript.
HBB created the analysis for the background due to scattered events and the corresponding plots, and edited the manuscript.
MN and AAST supervised the research and edited the manuscript.
All authors read and approved the final manuscript.
\subsection*{Acknowledgments}
The authors acknowledge Dr Gary Smith for the original funding application for this project, although he was unable to be involved in the research.


\bibliography{sn-bibliography}


\end{document}